\documentclass[12pt]{article}
\usepackage[koi8-r]{inputenc}
\usepackage{psfig}
\begin{document}
\begin{center}
{\LARGE \bf Infrared luminosities of galaxies in the Local Volume}
\bigskip

{\large I.D.Karachentsev$^1$  and   A.M.Kut'kin$^2$}
\bigskip

$^1$  Special Astrophysical Observatory, Russian Academy of Sciences,
Russia

$^2$  Moscow State University, Russia
\end{center}

\begin{abstract}
   Near-infrared properties of 451 galaxies with distances $D \leq$ 10
Mpc
are considered basing on the all-sky two micron survey (2MASS). A
luminosity function of the galaxies in the K-band is derived within
$[-25,-11]$ mag. The local ($D < 8$ Mpc) luminosity density is estimated
to be $6.8\times 10^8 L_{\odot}$/Mpc$^3$ that exceeds (1.5$\pm$0.1)
times the
global cosmic density in the $K$-band. Virial mass-to-K-luminosity
ratios are determined for nearby groups and clusters. In the luminosity
range of ($5\times 10^{10} - 2\times 10^{13})L_{\odot}$, the groups and
clusters follow
the relation  $\lg(M/L_K) \propto (0.27\pm0.03) \lg(L_K)$ with a scatter
of
$\sim$0.1 comparable to errors of the observables. The mean ratio
$<M/L_K> \simeq (20-25) M_{\odot}/L_{\odot}$ for the galaxy systems
turns out
to be significantly lower than the global ratio,
$(80-90)M_{\odot}/L_{\odot}$,
expected in the standard cosmological model with the matter density of
$\Omega_m =0.27$. This discrepancy can be resolved if most of dark
matter in the universe is not associated with galaxies and their
systems.
\end{abstract}

\section{Introduction}

 Recently published catalog of nearby galaxies (Karachentsev et al.
2004)
contains basic observational data on 451 galaxies with distances within
10 Mpc. Along with distances and radial velocities it contains different
optical characteristics of galaxies: apparent magnitude in B-band,
angular
size, morphological type, and also neutral hydrogen line flux and width
of
HI line. Data presented in the catalog allow one to determine such
important
properties of the Local Volume as luminosity function, local luminosity
density, mass density of baryonic matter or of neutral hydrogen.
Being marginally affected by selectional effects, distance (not flux)
limited sample of galaxies is the proper standard for comparison with
distant samples (like Deep Hubble Field) in analyzing effects of
galaxies
evolution.

   However detailed research of evolutional effects needs in photometric
data on the Local Volume not only in B-band, but also in other bands,
especially in near Infrared. There are two main reasons for this:
1) variable luminosity of young stellar population of galaxies, that
is increasing with star formation rate, introduce small contribution
in integral infrared (IR) luminosity of galaxy; 2) decreasing of
galaxy's
IR-luminosity due to light extinction in galaxy's dusty clouds is
far less than in B-band. That is why galaxy's IR luminosity is more
reliable indicator of baryonic mass than optical luminosity.

  In recent years all-sky survey in IR-bands ( J(1.1 -1.36) microns,
H(1.50-1.80) microns, and Ks(2.00-2.32) microns) called 2MASS survey was
conducted.
This survey became the base for extended objects survey, XSC, that
contains near 1.65 millions galaxies with apparent magnitudes in
Ks-band $<$ 14 mag and angular sizes $> 10$ arcsec (Cutri et al., 1998,
Jarrett,
2000). Different photometric and morphological properties of that huge
galaxies sample is described by Jarrett et al.(2000, 2003). We used
the 2MASS survey data for determination J, H, K magnitudes of galaxies
in the Local Volume.

\section{ Galaxies in the Local Volume, visible in the 2MASS survey.}

Infrared J, H, K images of all galaxies with distance estimates
within 10 Mpc were examined with NASA Extragalactic Database (NED).
After excluding of some cases with questionable identifications of
distant 2MASS sources with a nearby galaxy we obtained the sample of 122
galaxies for which the 2MASS XSC catalog contains J, H, K- "extended"
magnitudes. Unlike measured (isophot) magnitudes, J ,H, K extended
magnitudes take into account contribution of faint external parts
of galaxy by extrapolation along it's standard brightness profile.
Thus, only 27\% of the Local Volume galaxies are seen in the 2MASS
survey.

The reason for absence of many nearby galaxies is relatively short
exposition time ($\sim$8 sec per frame). And, as a consequence, galaxies
of low surface brightness fall below the survey sensitivity threshold,
K = 20 mag/sq.arcsec. The low detection rate in the 2MASS survey is
typical
indeed for galaxy samples limited by distance. For instance, the spiral
galaxies
of RFGC catalog with angular diameters larger than 0.6 arcmin have a
detection rate in the 2MASS about 71\%
(Karachentsev et al., 2002). Appearance of a galaxy in the 2MASS survey
depends on it's different characteristics: morphological type, absolute
magnitude, mean surface brightness. To estimate the J, H, K magnitudes
for
galaxies of the Local Volume, that are not seen in the 2MASS, we built
dependence of color indexes B-K, J-K, H-K for detected galaxies on
their morphological types and other parameters. Distribution of 122
nearby galaxies on color index B-K and morphological type in de
Vaucouleur's
classification scheme is resulted in upper side of plot 1. Each galaxy
is shown as a cross and the median values of color index B-K are shown
as
diamonds. All apparent magnitudes were corrected for galaxy extinction:
$A_J=0.209A_B , A_H=0.133A_B,$ and $ A_K=0.085A_B$, where
$A_B$ is the magnitude of extinction in the B-band from
Schlegel et al. (1998). These data  show a systematic trend of the mean
color index from $<B-K> \simeq 4.0$ for early types to $<B-K> \simeq
2.0-2.5$ for
later classes. Similar tendency is also visible for color indexes J-K
(lower side in plot 1). The obtained
color index dependences on galaxies type satisfactorily match
data presented in Jarrett et al. (2003) on plot 20 and 23 for more
numerous sample of galaxies of different types. Following Jarrett
et al. (2003) correlations "type - color index" were used for
determination of J, H, K-magnitudes for nearby galaxies from
their B-magnitudes contained in the catalog (Karachentsev et al. 2004).
Results are presented in Appendix accessible as an electronic file.
The most useable names of the nearby galaxies are given in the same
sequence as in the parent catalog. The infrared magnitudes from 2MASS
are shown with two signs after dot, and the calculated magnitudes J, H,
K
are presented with one sign after dot. Comparison of different methods
in estimation of the J, H, K magnitudes showed that a typical error in
IR-magnitudes
determination is about 0.5 mag.

\section{Luminosity function of galaxies in the K-band.}

We determined absolute magnitudes of all galaxies from the Local
Volume (LV) basing on J, H, K values from Appendix, using galaxy
distance estimates from the catalog (Karachentsev et al., 2004) and
taking into account Galactic extinction from Schlegel et al.
(1998). Plot 2 shows distribution of sample of the nearest galaxies
along absolute magnitudes in J (crosses), H (diamands), and K
(circles) bands. The most bright galaxy in the LV,
NGC 4594 ("Sombrerro") has absolute magnitude $M_K=-24.91$, and
the most faint dwarf systems of the Ursa Minoris type have $M_K
\sim-11$.
The maximum of the luminosity function in all J, H, K-bands falls on
interval [-15, -17].

Recently Cole et al. (2001) made an attempt
to determine galaxy's luminosity functions in J and K band.
They used the 2MASS photometry of 17173 galaxies with
measured redshifts from 2dF survey. Accordingly Cole et al. (2001),
luminosity function in the K-band is well represented by Shechter's
function  with parameters: $M^* = -24.15 (H_0 = 72$ km/s Mpc) and
$\alpha = -0.96$ on an interval from $-26^m$ to $-18^m$. Luminosity
function
comparison for the 2MASS/2dF sample (squares) and the LV
(solid circles) sample is presented in upper side of Plot 3.
Statistical errors, $(N)^{1/2}$ are shown as vertical bars.
Mutual normalization of two the samples is done near $M_K = -20.5^m$,
where statistical errors for these samples are comparable.
The luminosity function for the LV galaxies expectedly
reaches far away (on seven magnitudes) toward dwarf galaxies.  In
overlapping region, for $M_K = [-25^m, -18^m]$, consistency of
luminosity functions can be treated as satisfactory, although
differences run up to 2-3 standard deviations for some bins.
There's a remarkable flattering of the 2dF luminosity function
in $M_K = [-19^m, -18^m]$ region. The obvious reason is
because of systematic incompleteness in the 2MASS/2dF sample, so
there are many low luminosity dwarf galaxies with low surface
brightness that are not detected in the 2MASS survey. That is why
there's a much better consistency of luminosity function for
the LV based on only 122 detected in the 2MASS objects
(crosses) with Cole's sample (Cole et al., 2001).

Another possibility for comparison is data
on the Virgo cluster of galaxies that can be found in Internet on
http://goldmine.mib.infn.it/. This sample contains K-magnitudes
for bright members of the Virgo cluster. We estimated K-magnitudes
of fainter members being based on their B-magnitudes and dependency
of mean color index $< B - K >$ on morphological type that was
found in Jarrett et al. (2003). As a result, the luminosity function
for 680 members of the Virgo cluster was obtained (triangles in low
side of Plot 3). Mutual normalization of distributions for the Virgo
cluster, the LV sample (circles) and the
2MASS/2dF sample (squares) was made for $M_K = -20.5$. The
distance to the Virgo cluster was taken equal to 17.0 Mpc (Tonry
et al. 2000). As can be seen, the Virgo cluster luminosity function
reveals some excess in supergiant galaxies that is expected
in terms of formation scenario for the most massive galaxies as
a merging result for normal and dwarf galaxies.
On the faint end it extends up to $M_K = -13^m$, that is on
5 magnitudes further than for the 2MASS/2dF sample. Systematic
differences between the Virgo cluster and the LV samples on
the faintest absolute magnitudes are obviously due to underestimation
in number of cluster's dwarf members, for which estimations of radial
velocities are still missing. The last fact doesn't allow one to
either rank them as members of the Virgo cluster or as just background
galaxies.

Using distribution of the LV galaxies
on absolute magnitudes and distances we calculated the mean luminosity
density as a function of sphere radius centered in our Galaxy.
Plot 4 shows a luminosity density (in sun luminosity per cubic Mpc
units)
dependence on distance separately for J, H and K-bands. The mean
luminosity
density falls down on two orders of magnitude on scales from 1 to 10
Mpc, reflecting fractal character of galaxies distribution.
Completeness of the catalog was estimated as on  70\%  level within
distance D=8 Mpc. Some dwarf galaxies in outlying regions of the
LV $(D \simeq 5 - 10$ Mpc) and in regions of strong extinction on
galactic latitudes I$b$I $\leq 10^{\circ}$ could be easly missed from
the
catalog.
However lack of dwarf galaxies in practice doesn't affect on estimation
of luminosity density because more than 90\% of the integrated
luminosity
accounts on bright galaxies constituting interval that is within 4
magnitudes from the level of the brightest galaxy.

       Photometry of galaxies in the 2MASS survey was
used by Kochanek et al. (2001) and Bell et al. (2003) for global
mean luminosity density estimation in K-band, $j_K$. Using data on
redshifts for galaxies from the Sloan sky survey and other sources,
these authors obtained values $j_K = (5.1\pm0.5)10^8
L_{\odot}$/Mpc$^3$ and
$(4.2\pm1.3)\times 10^8L_{\odot}$/ Mpc$^3$, correspondingly. These
values are shown
as two horizontal lines on Plot 4. Thus, our calculated K-band
mean luminosity density within 8 Mpc, $6.8\times10^8L_{\odot}$/ Mpc$^3$,
 (1.4-1.6)
times exceeds the global one. Note that a ratio of local to
global mean luminosity density in the B - band, that is (1.4 - 2.0),
contains
more uncertainty because of different methods for
accounting of internal absorption in galaxies.

\section{Mass-to-K-luminosity ratio for nearby galaxies.}

   Almost 2/3 of the LV galaxies were detected in the HI line.
That allowed us to determine mass of hydrogen in these galaxies,
M(HI), using observed flux in HI, as well as the total galaxy mass
inside the standard radius, $M_{25}$, using a width of the HI line
corrected for the galaxy inclination. The total mass-to-luminosity
ratio and the HI mass-to-luminosity ratio are important global
parameters
of a galaxy that depend on the features of galaxy evolution: star
formation rate, frequency of merges etc. Mass-to-blue luminosity ratio
distribution for the LV galaxies was discussed by Karachentsev et
al.(2004).
Distribution of $M_{25}/L$ calculated for the B and
the K-bands for the nearby galaxies is shown in the upper side of
Plot 5. Similar ratio for hydrogen mass per unit luminosity in
the B and the K-bands is shown in the lower side of Plot 5. Galaxies
with
directly determined K magnitudes from the 2MASS survey are shown as
squares
and galaxies with estimations of IR-luminosities through their color
(B - K) and morphological type shown as crosses. As can be seen
the transition from optical to infrared luminosities of galaxies doesn't
lead to expected decreasing in dispersions for $M_{25}/L$ and M(HI)/L
values. This obviously results from a low accuracy in determination
of IR-luminosity for nearby galaxies, which are mostly low surface
brightness dwarf galaxies.

Distribution of integrated K-luminosity and $M_{25}/L_K$ ratio for the
LV
galaxies is shown in logarithmic scale in Plot 6. It is evident from
these
data that the mean mass-to-luminosity ratio remains nearly constant
over the IR-luminosity range of 5 orders being equal to
$(1.5\pm0.2) M_{\odot}/L_{\odot}$.
Basing on Salpeter's conception on initial stellar mass
function, Persic\& Salucci (1992), Fukugita et al. (1996),
Kochanek et al. (2001), and Cole et al. (2001) estimated the mean
stellar
mass density in the Universe. Following them and Bell et al. (2003), the
stellar component contains $\Omega_* = (2.8\pm0.8)\times10^{-3}$ from
the
critical density
with H=72 km/s Mpc or $(M/{L_K})_* = 0.95\pm0.27$ in solar mass and
luminosity
units. Thus, the stellar mass component (+ mass of gas) consistent
within
error range with the mean galaxy mass inside it's standard radius.
Consequently, nonvisible forms of matter (dark halo) contribute
secondary in the integrated mass of galaxy inside it's standard radius.
This is considered both to normal and dwarf galaxies within $10^{11}$
to $10^6 L_{\odot}$ luminosity range.

\section{ Mass-to-K-luminosity ratio in systems of galaxies}

      As well known, mass-to-luminosity ratio in
systems of galaxies increases with transition from double and triple
systems to groups, clusters and superclusters. Character of this
dependency in the B- band was discussed by Karachentsev (1966),
Bahcall et al. (2000) and many others. There are reasons to believe
that this dependence must be clearer in IR then in optical.
Recently Lin et al. (2003) explored IR properties of 27 galaxy clusters
and showed that mass dependence on K-luminosity of clusters expresses
in relation $M/L_K\propto M^{0.31\pm0.09}$, where masses of clusters
were
obtained
using their X-ray fluxes. Estimations of $M/L_K$ were made for some
others
systems of galaxies too. Summary of these data can be found in Table 1,
where we also show masses and K-luminosities obtained for nearby groups
of galaxies. First column indicates name of galaxy or system, in some
cases we give the spatial scale using in the mass determination. Second
column indicates integrated K-band luminosity of the object/system in
units
of solar luminosity. Third and fourth columns
indicate total mass and mass-to-K-luminosity ratio, respectively.
And the last one contains information about sources of the mass
estimations.
\begin{table}[h]
\caption{Mass-to-luminosity ratios for galaxies and systems of galaxies
in the K-band.}

\begin{tabular}{lccll} \hline
 Object           &    $L_K$   &     $M$    &   $M/L_K$     & Notes
	 \\
		  &  $(L_{\odot})$ &  $(M_{\odot})$ &   $(\odot)$    &
		  \\
\hline
LV galaxies    &  1.0 E07 &  1.5 E07 &  1.5$\pm$0.2  &  $M_{25}/L$
	\\
LV galaxies    &  1.0 E09 &  1.3 E09 &  1.3$\pm$0.2  &  $M_{25}/L$
	\\
LV galaxies    &  1.0 E11 &  1.1 E11 &  1.1$\pm$0.2  &  $M_{25}/L$
	\\
NGC 5128 ($<$80kpc)   &  1.1 E11 &  5.0 E11 &  4.7       &  Peng et
al.(2004)
   \\
NGC 4636 ($<$35kpc)   &  1.6 E11 &  1.5 E12 &  9.7       &  Loewenstein
\& Mushotsky (2002)  \\
NGC 1399 ($<$106kpc)  &  2.8 E11 &  5.2 E12 & 18.7$\pm$5.7  &  Jones et
al.(1997)
      \\
M31 group        &  6.3 E10 &  8.4 E11 & 13.4       &  Karachentsev
(2005)
 \\
Local Group    &  1.1 E11 &  1.2 E12 & 11.2$\pm$3.5  &  Karachentsev
(2005)
    \\
 M81 group       &  1.5 E11 &  1.6 E12 & 10.6       &  Karachentsev
(2005)
  \\
IC 342 group      &  5.6 E10 &  7.6 E11 & 13.5       &  Karachentsev
(2005)
  \\
 Maffei group     &  7.2 E10 &  1.0 E12 & 14.0       &  Karachentsev
(2005)
  \\
M83 group       &  6.8 E10 &  1.0 E12 & 15.2       &  Karachentsev
(2005)
  \\
 CenA group      &  1.4 E11 &  3.0 E12 & 21.3       &  Karachentsev
(2005)
  \\
 Leo-I group     &  3.5 E11 &  7.2 E12 & 20.5       &  Karachentsev \&
Karachentseva (2004)             \\
 NGC 6946 group     &  6.8 E10 &  8.0 E11 & 11.7       &  Karachentsev
et al.(2000) \\
Poor groups     &  7.6 E10 &  1.3 E12 & 17.0$\pm$2.9  &  Guzik \& Seljak
(2002)       \\
Fornax            &  1.8 E12 &  5.9 E13 & 32         &  Desai et
al.(2003)
   \\
Virgo ($<$1.6Mpc)   &  8.8 E12 &  4.2 E14 & 48$\pm$6      &  McLaughlin
(1999)
	\\
0024+1654         &  1.5 E13 &  6.1 E14 & 40$\pm$7      &  Kneib et
al.(2003)
      \\
MS0302+17 ($<$8Mpc) &  8.9 E12 &  4.4 E14 & 50$\pm$5      &  Gavazzi et
al.(2004)       \\
Abell clusters  &  1.1 E13 &  5.2 E14 & 47$\pm$3      &  Lin et
al.(2003)
       \\
Coma  ($<$14Mpc)    &  2.7 E13 &  1.4 E15 & 54$\pm$17     &  Rines et
al.(2001)
      \\
&  & & & \\
\hline
\end{tabular}
\end{table}

All data in Table 1 correspond to the Hubble constant $H_0 = 72$ km/s
Mpc
and the
solar absolute magnitude $M_{K,\odot}$ = 3.39 mag (Kochanek et al.,
2001).

It is not out of place to make some comments on values given in Table 1.
First three lines represent the mean values of masses inside the
standard radius for the LV galaxies, divided in three luminosity
intervals. The mean ratios $M_{25}/L_K$ and their standard deviations
are
derived
from data in Plot 6. Next three lines contains estimations of total
masses
for individual galaxies on larger scales. They were made using
dispersions of radial
velocities of planetary nebulas in elliptical galaxy NGC 5128 (on scale
of 80 Kpc) or
using value of X-ray flux from ellipticals NGC 4636 (in the Virgo
cluster) and
NGC 1399 (in the Fornax cluster). The K-values for these galaxies were
obtained in
the 2MASS.

Next seven lines contain data on masses and luminosities
for the nearest groups around the giant galaxies: M31, M81,
IC 342, Maffei 1, M83 and Cen A = NGC 5128 as well as for the Local
Group. Population and dynamics of these groups were discussed
by Karachentsev (2005). Presented masses of these group correspond to
the mean of two estimates: one made using Virial theorem and other one
basing on orbital motions of companions around their principal galaxies.
The value of total mass of the Local Group as a whole was obtained from
relation $M_t=(\pi^2/8G)\times R^3_0/T^2_0$, where $R_0$ is the
observed radius
of "the zero velocity
sphere", $T_0$ is the age of the Universe, and G is the gravitational
constant.
Another two groups of the LV were added to those listed above:
Leo-I and NGC 6946. Their virial masses were estimated by Karachentsev
\& Karachentseva (2004) and Karachentsev et al. (2000), respectively,
with
additional data on radial velocities from Makarov et al. (2003).

Guzik \& Seljak (2002) and Hoekstra et al. (2004)
explored effects of weak gravitational lensing that produced by single
galaxies of high luminosities seen in SDSS and RCS surveys. Such
galaxies
are usually inhabit centers of pure populated (loose) groups.
Accordingly authors, lensing galaxies with the mean optical luminosity
$L_B=1.9\times10^{10}L_{\odot}$ have characteristic masses
($1.3\pm0.2)
\times10^{12} M_{\odot}$ on
scales $R\leq 260$ Kpc. Supposing the mean color for these field
galaxies
to be $< B - K > = 3.5$, we obtain the estimation of
mass-to-K-luminosity
ratio ($17\pm3)M_{\odot}/L_{\odot}$ that is in good agreement with the
data on
nearby
groups of galaxies mentioned above.

Last six lines in Table 1 refer to clusters and superclusters.
According to Tonry et al. (2001) and Jerjen (2003) the distance to pure
populated southern cluster in Fornax is $20\pm$2 Mpc.  A catalog
of 2678 galaxies from that cluster was published by Ferguson (1989).
With a velocity dispersion of the Fornax members being $\sim$400 km/s,
the
virial
mass of that cluster is $5.9\times 10^{13} M_{\odot}$ (Desai et al.
2004). The
infrared
luminosity of the Fornax cluster presented in Table 1 was obtained by us
from the 2MASS survey data.

Analogous method was used to estimate the integrated K-luminosity
for other nearest galaxy cluster in Virgo. With the Virgo cluster
distance
being equal D = 17 Mpc, we obtained
$L_K$(Virgo)$=8.8\times10^{12}L_{\odot}$
for the virialised
region of the cluster within 6 deg from the center. The virial mass of
the cluster
was obtained by McLanghlin (1999) from a velocity dispersion of galaxies
and
from an X-ray flux of virialized hot gas within cluster.  The mass of
the Vigro
is $(4.2\pm0.5) 10^{14} M_{\odot}$ within 1.6 Mpc radius. That gives
$M/L_K
=(48\pm6)M_{\odot}/L_{\odot}$. It is worth to say that according to a
model of
galaxy
motions around the Virgo as an attractor, Tully \& Shaya (1984) and
Tonry et al. (2000) obtained mass of the cluster $7\times 10^{14}
M_{\odot}$.
However
this estimate concerns larger cluster radius up to $\sim$8 Mpc, and
the integrated K-luminosity of the volume is not determined so far.

Recently Kneib et al. (2003) and Gavazzi et al. (2004) used effects of
gravitational lensing for mass determination of two rich clusters:
0024+1654 and MS0302+17. Their results are presentin Table 1.
The authors measured K-luminosity for first cluster. Estimation of
K-luminosity for second one was made using known luminosity in B-band
in suggestion that cluster is inhabited by galaxies of early types
with the mean color index $< B - K >$ = 4.0.

Two the last lines in Table present the mean mass and
K-luminosity for 27 clusters that were studied by Lin et al.
(2003) and also for the Coma supercluster (Rines et al. 2001). In
last case $M/L_K$ ratio was measured using redshift survey of 1779
galaxies
and their 2MASS photometry. Authors point that obtained
mass-to-luminosity
ratio remains nearly constant while move out from central virialized
region $\sim$3.5 Mpc in radius to a scale $\sim$14 Mpc, that
characterizes a
region
of bulk motion for surrounding galaxies toward Coma as an attractor.

Data on masses and K-luminosities of galaxies and systems of galaxies
summarized in Table 1 are presented in Figure 7. The estimates of M/L
for individual galaxies, groups, and clusters/superclusters
are shown as circles, triangles and squares, respectively. We can get
the
global value $(M/L_K)_{global}$ that characterizes fairish (on scales
$>$
100 Mpc)
regions of the Universe calculating ratio of critical density
$\rho_c = (3H_0^2/8\pi G$) (that equals $1.43\times 10^{11}M_{\odot}$
Mpc$^{-3}$
for $H_0 = 72$ km/s Mpc) to mean luminosity density in
the K-band, $j_K$ .Using estimations of $j_K$ from Kochanek et al.(2001)
and Bell et al. (2003) we obtain $(M/L_K)_{global} = 287$ and $345
M_{\odot}/L_{\odot}$,
respectively. These values are shown as dashed lines in
right upper corner of Figure 7. We accepted their mean
$(316\pm29)M_{\odot}/L_{\odot}$
for calibration of density scale in units of the critical density. The
last is shown on the right side.

The mean stellar mass density,  $(0.95\pm0.27)M_{\odot}/L_{\odot}$
from Bell et al. (2003), is shown at the bottom of the
Figure 7 as horizontal line. Other horizontal line corresponds to
the mean baryonic mass density, $\Omega_b = 0.047\pm0.006$ or
$<M/L_K>_b=
(14.8\pm1.9)M_{\odot}/L_{\odot}$ according to Spergel et al. (2003).
Dashed
line with a weak slope corresponds to a trend of the mean mass (inside
standard radius) to luminosity ratio for the LV galaxies.
Analysis of the data presented allows one to make some general
conclusions:

1) The value of stellar mass density and gas density per unit
luminosity is quite enough to explain the mean mass-to-luminosity ratio
inside the standard radius both for giant, normal and dwarf galaxies.
With all this going on, contribution of gaseous component, that can
be expressed through relation $M(HI)/M_{25}$, increases with decreasing
galaxy's luminosity, reaching values $\sim$(1-3) for the smallest dwarf
systems (compare Fig. 10 in Karachentsev et al., 2004)

2) Transition from visible regions of galaxies with sizes (1-30) Kpc
to groups dominated by one galaxy (scales 50-300 Kpc) is accompanied by
increasing of mass on near order of magnitude with almost constant
luminosity.
The mean ratio $M/L_K$ for groups of galaxies is actually coincide with
the mean ratio for baryonic mass, $(14.8\pm1.9)M_{\odot}/L_{\odot}$.

3) Galaxy groups, clusters and superclusters follow a relation
$\lg(M/L_K)\propto
(0.27\pm0.03)\lg L_K$ within luminosities interval from $5\times10^{10}$
to
$2\times10^{13}L_{\odot}$.
This relation has a typical dispersion $\sim$0.1 that is comparable with
uncertainties in measuring of integral luminosities and masses. At the
same time mutual differences in mass estimations based on virial
motions,
X-ray flux or lensing turn out to be inessential.

4) The largest and the most inhabited systems of galaxies --- rich
clusters/superclusters are characterized by values of mass-to-luminosity
ratios being $M/L_K\simeq 50 M_{\odot}/L_{\odot}$, that constitutes only
16\% from
the global
value $(M/L_K)_{global} = (316\pm29)M_{\odot}/L_{\odot}$. Since only
small
part of galaxies
($\sim$10-20\% constitute to such supersystems, an ensemble averaged
$<M/L_K>\simeq(20-25)M_{\odot}/L_{\odot}$ turns out to be certainly
lower than
that  expected
$<M/L_K>_m = (80-90)M_{\odot}/L_{\odot}$ in the standard cosmological
model
(Spergel et
al. 2003) with $\Omega_m = 0.27$ and $\Omega_{\lambda} = 0.73$. To
adjust
together
the ensemble averaged value $\Omega_{DM}\sim 0.07$ with the global value
$\Omega_m = 0.27$
one needs to suppose that most part of dark matter in the Universe is
not
associated with galaxies or their systems (Karachentsev, 2005).

\section{ Concluding remark.}

As can be seen from the results above, IR-luminosities of nearby
galaxies
play important role in exploring dynamics and evolution of galaxy
systems. Realization of the all-sky survey in J, H, K-bands and creation
of the XSC 2MASS catalog allow one to investigate different properties
for
big samples of galaxies in unified photometric system. Determination  of
IR
characteristics for a sample of nearby galaxies, which is distance, but
not
flux-limited, is an important task. Parameters of such sample must be
marginally affected by different selectional effects, which complicate
interpretation of initial data from observations. Unfortunately, among
450 galaxies of the Local volume within D = 10 Mpc only a small part
(27\%) is
seen in the 2MASS because of low luminosity, low surface brightness or
blue
color. That is why measuring J, H, K-magnitudes for all galaxies inside
the LV for the purpose of formation of reference sample seems to be a
very
promising aim. This sample could be an etalon while comparing it with
deeper samples on different redshifts.

 {\bf Acknowledgements.}  This work was partially supported by RFBR
grant
04-02-16115
and DFG-RFBR grant 02-02-04012.

{\bf References}

Bahcall N.A., Cen R., Dave R. et al., 2000, Astrophys. J.,541, 1.

Bell E.F., McIntosh D.H., Katz N., Weinberg M.D., 2003, Astrophys. J.
Suppl.,149, 289.

Cole S., Norberg P., Baugh C.M., et al., 2001, Monthly Not. Roy. Astron.
Soc., 326, 255.

Cutri R.M., Skrutskie M.F., 1998, Bull. Am. Astron. Soc., 30, 1374.

Desai V., Dalcanton J.J., Mayer L., et al., 2004, Monthly Not. Roy.
Astron. Soc., 351, 265.

Ferguson H.C., 1989, Astron. J., 98, 367.

Fukugita M., Ichikawa T., Gunn J., et al., 1996, Astron. J., 111, 1748.

Gavazzi R., Mellier Y., Fort B., et al., 2004, Astron. and Astrophys.,
422, 407.

Guzik J., Seljak U., 2002, Monthly Not. Roy. Astron. Soc., 335, 311.

Hoekstra H., Yee H.K., Gladders M.D., 2004, Astrophys. J., 606, 67.

Jarrett T.H., 2000, Publ. Astr. Soc. Pacific, 112, 1008.

Jarrett T.H., Chester T., Cutri R., et al., 2000, Astron. J., 119, 2498.

Jarrett T.H., Chester T., Cutri R., et al., 2003, Astron. J. 125, 525.

Jerjen H., 2003, Astron. and Astrophys., 398, 63.

Jones C., Stern C., Forman W. et al., 1997, Astrophys. J., 482, 143.

Karachentsev I.D., 1966, Astrofizika, 2, 81

Karachentsev I.D., 2005, Astron. J., 129, N1 (accepted).

Karachentsev I.D., Karachentseva V.E., 2004, Astron. Zh., 81, 298.

Karachentsev I.D., Karachentseva V.E., Huchtmeier W.K., Makarov D.I.,
2004, A Catalog of Neighboring Galaxies, Astron. J., 127, 2031

Karachentsev, I.D., Mitronova S.N., Karachentseva V.E., et al., 2002,
Astron. and Astrophys., 396, 431.

Karachentsev I.D., Sharina M.E., Huchtmeier W.K., 2000, Astron. and
Astrophys., 362, 544.

Kneib J.P., Hudelot P., Ellis R.S., et al., 2003, Astrophys. J., 598,
804.

Kochanek C.S., Pahre M.A., Falco E.E. et al., 2001, Astrophys. J., 560,
566.

Lin Y.T., Mohr J.J., Stanford S.A., 2003, Astrophys. J., 582, 574.

Loewenstein M., Mushotzky R., 2002, astro-ph/0208090.

Makarov D.I., Karachentsev I.D., Burenkov A.N., 2003, Astron. and
Astrophys., 405, 951.

McLaughlin D.E., 1999, Astrophys. J., 512, L9.

Peng E.W., Ford H.C., Freeman K.C., 2004, Astrophys. J., 602, 685.

Persic M., Salucci P., 1992, Monthly Not. Roy. Astron. Soc., 258, 14.

Rines K., Geller M.J., Kurtz M.J., et al., 2001, Astrophys. J., 561,
L41.

Schlegel, D.J., Finkbeiner, D.P., Davis, M., 1998, Astrophys. J., 500,
525.

Spergel D.N., et al., 2003, Astrophys. J. Suppl., 148, 175.

Tonry J.L., et al., 2000, Astrophys. J., 530, 625.

Tonry J.L., et al., 2001, Astropys. J., 546, 681.

Tully R.B., Shaya E.J., 1984, Astrophys. J., 281, 31.

\begin{figure}[hbt]
\vbox{\includegraphics{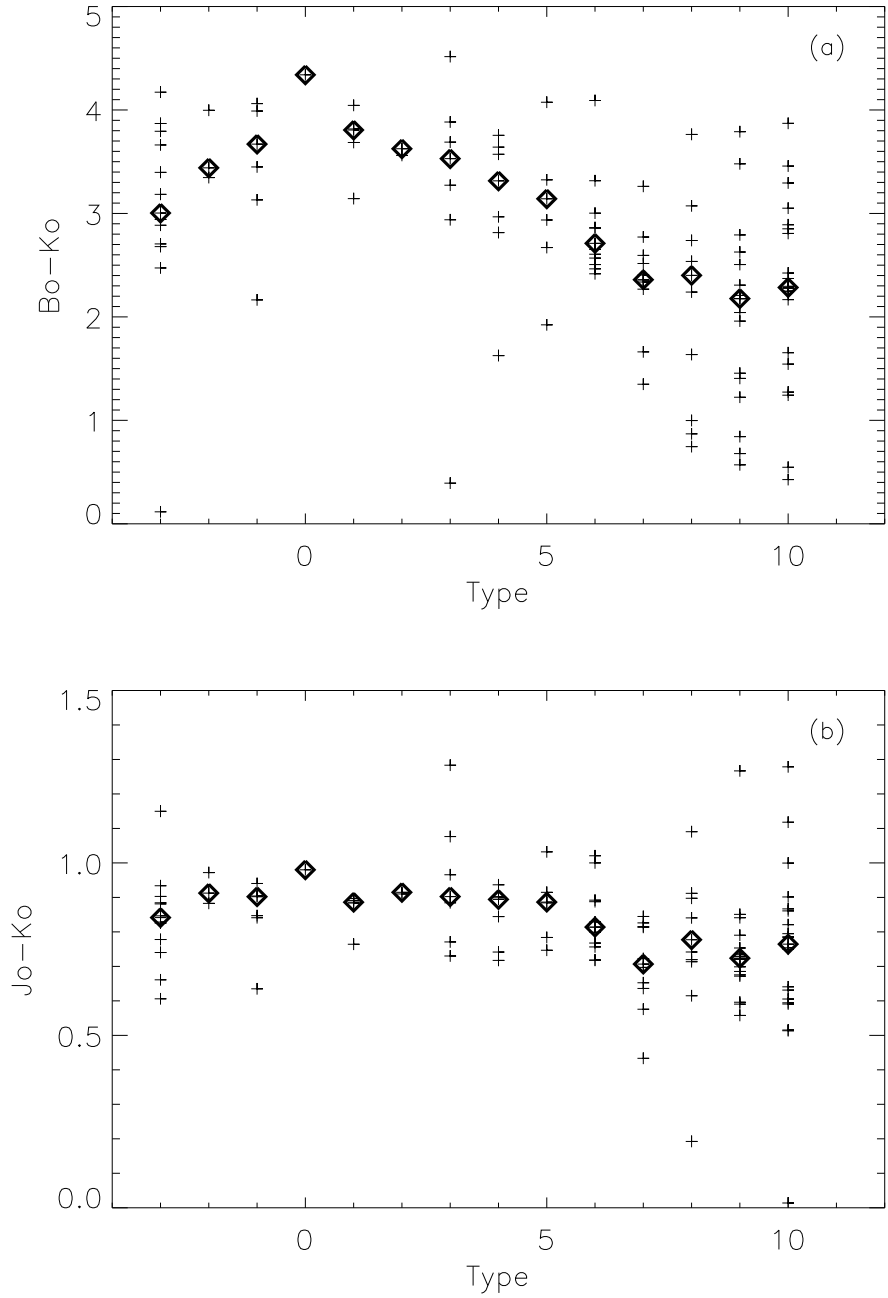}}\par
\vspace{15.0cm}
\caption{  B-K (a) and J-K (b) colors for 122 galaxies of different
morphological
       types in the Local Volume.}
\label{pic1.ps}
\end{figure}

\begin{figure}[hbt]
\vbox{\includegraphics{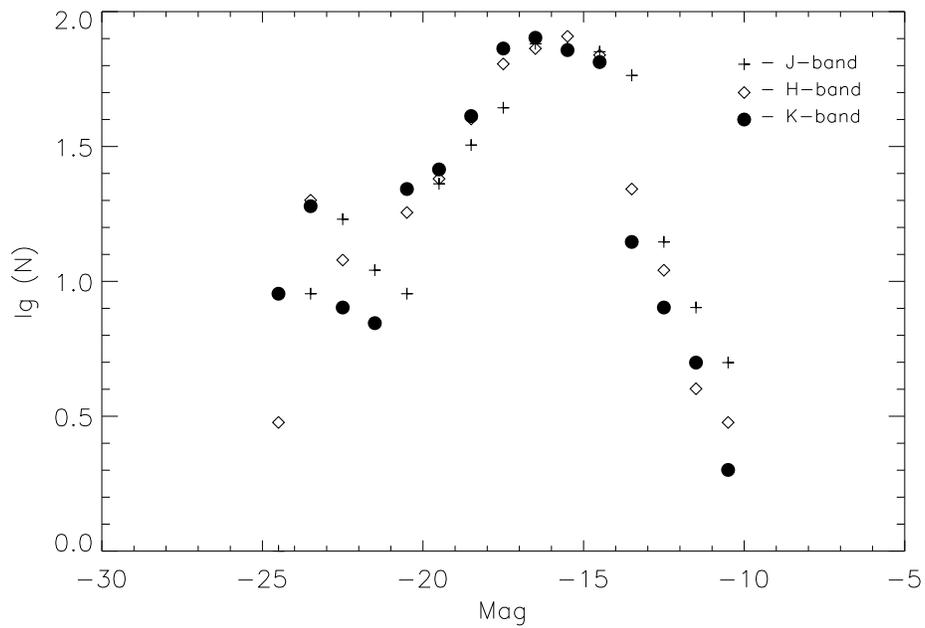}}\par
\vspace{7.0cm}
\caption{ Number distribution of 451 galaxies in the LV according to their
absolute
       magnitudes in the J, H, and K bands.}
\label{pic2.ps}
\end{figure}

\begin{figure}[hbt]
\vbox{\includegraphics{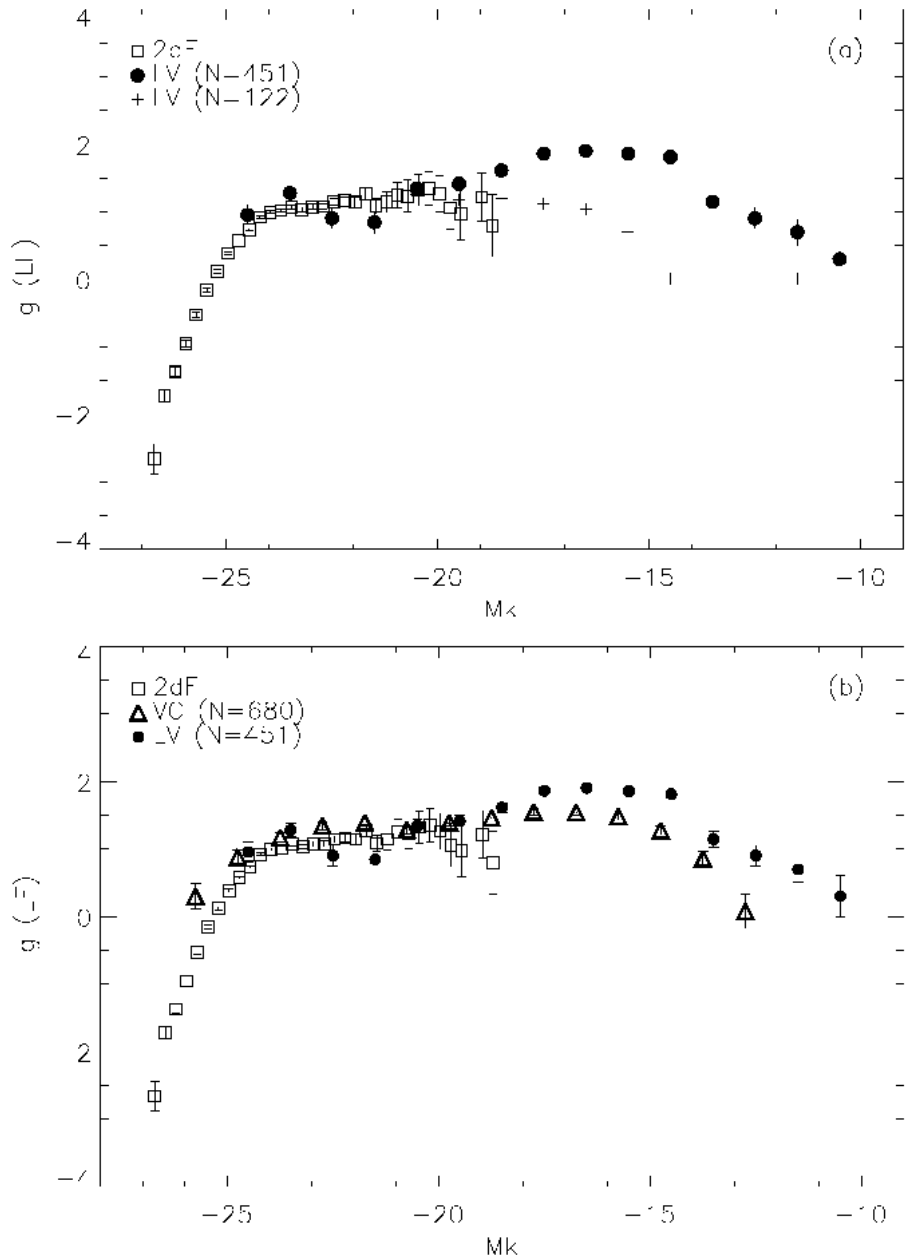}}\par
\vspace{15.0cm}
\caption{Luminosity function (LF) for 451 LV galaxies in the K-band (solid
circles)
       in comparison with the LF for galaxies from the 2dF survey
(squares)
       according to Cole et al. (2001). The LF for 122 LV galaxies
detected in
       the 2MASS is shown by crosses (a), and the LF for 680 members of
the Virgo
       cluster is shown by triangles (b).}
\label{pic3.ps}
\end{figure}

\begin{figure}[hbt]
\vbox{\includegraphics{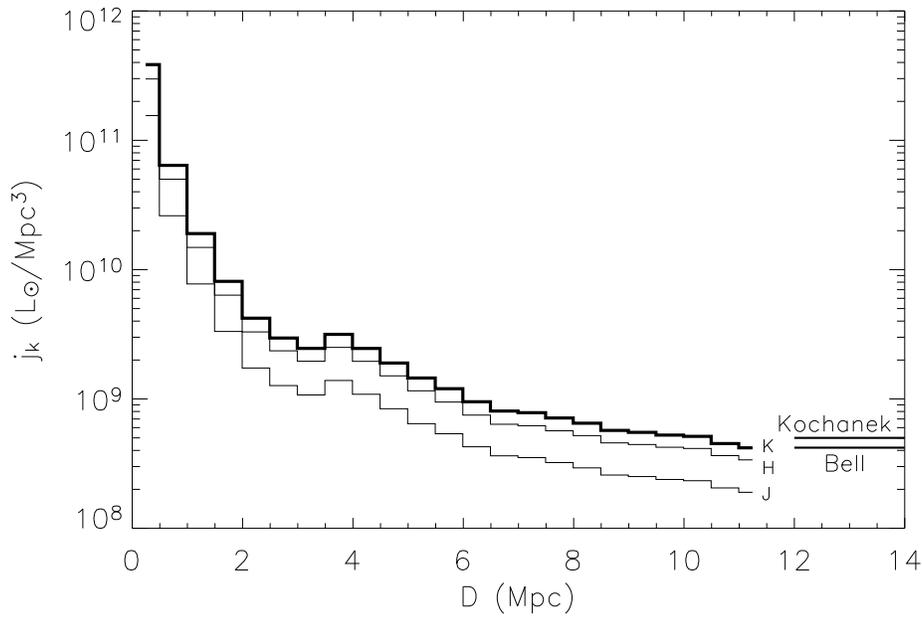}}\par
\vspace{7.0cm}
\caption{ The mean luminosity density of nearby galaxies in the J,H, and K
bands
       versus galaxy distance D in Mpc. The global luminosity density in
the
       K- band from Kochanek et al. (2001) and Bell et al. (2003) is
shown by
       two horizontal lines at right side.}
\label{pic4.ps}
\end{figure}

\begin{figure}[hbt]
\vbox{\includegraphics{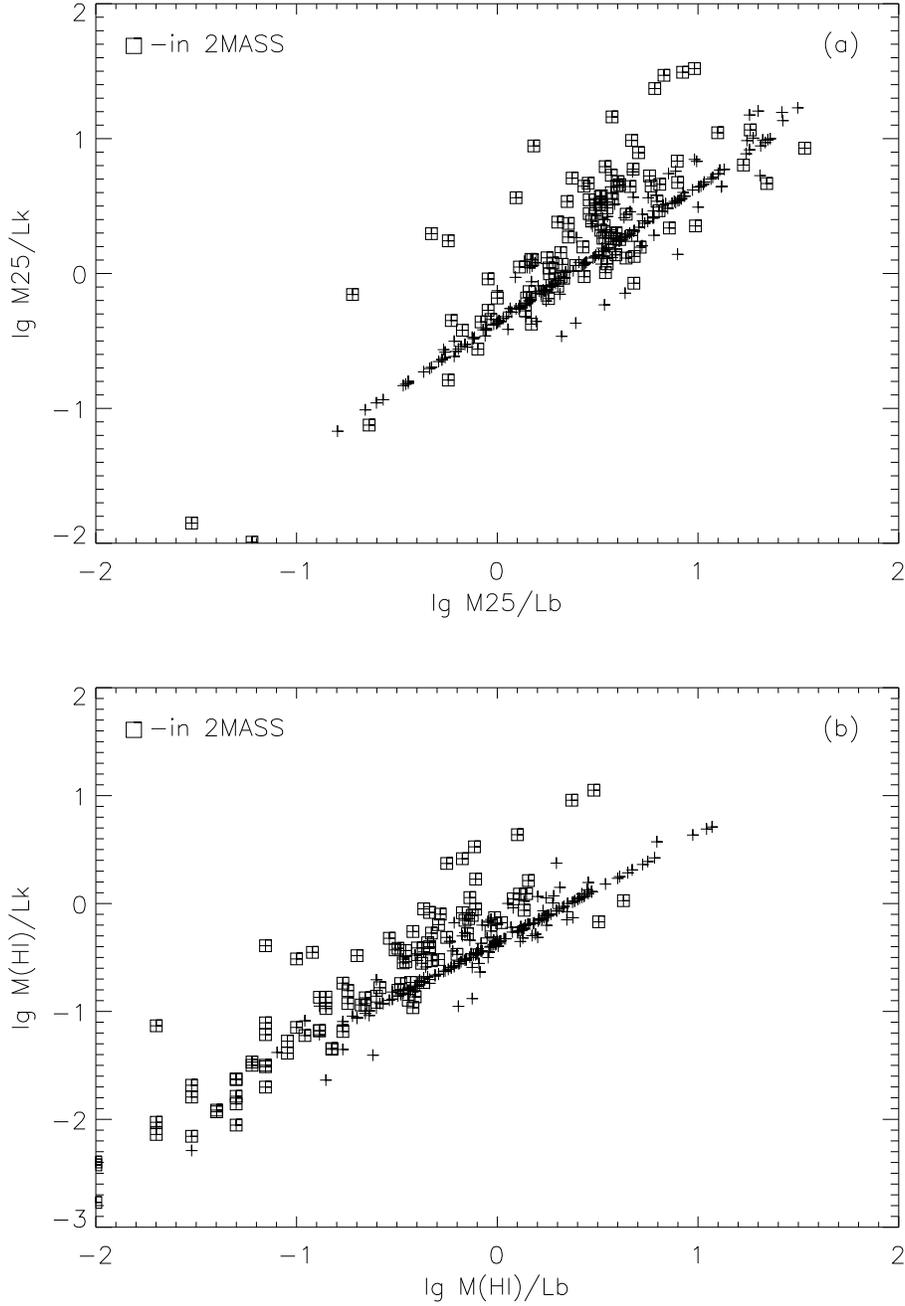}}\par
\vspace{15.0cm}
\caption{ Total mass (inside the standard radius, $R_{25}$) to luminosity
ratio (a)
       and hydrogen mass to luminosity ratio (b) for the LV galaxies in
the B and
       K bands. The galaxies detected in the 2MASS are shown by squares,
and
       objects with K-luminosity calculated from their mean B-K color
and
       morphological type are shown by crosses.}

\label{pic5.ps}
\end{figure}

\begin{figure}[hbt]
\vbox{\includegraphics{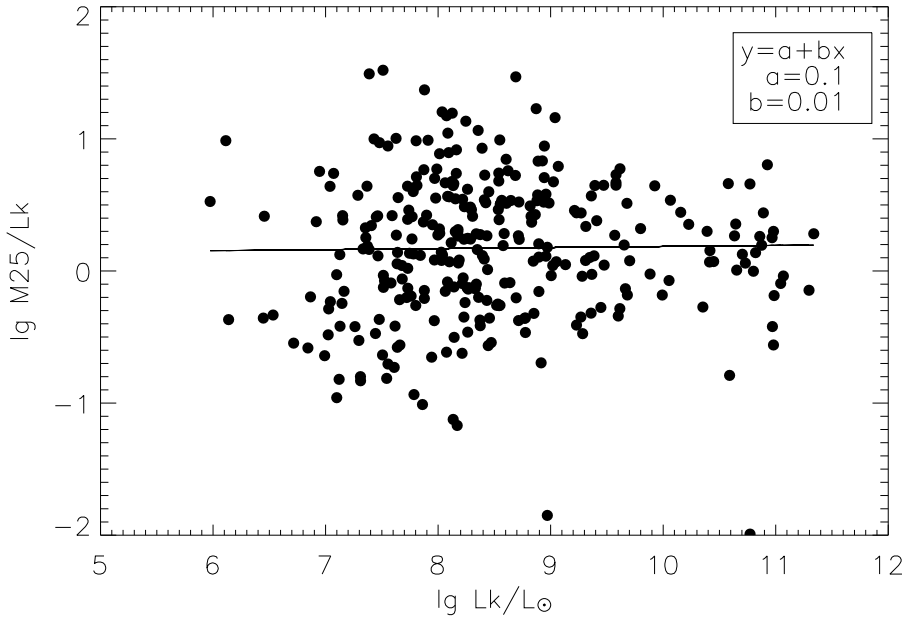}}\par
\vspace{5.0cm}
\caption{ Total mass to K-luminosity ratio for the LV galaxies versus their
       K-luminosity. The linear regression is shown by a stright line
with
       parameters indicated in the upper right corner.}
\label{pic6.ps}
\end{figure}

\begin{figure}[hbt]
\vbox{\includegraphics{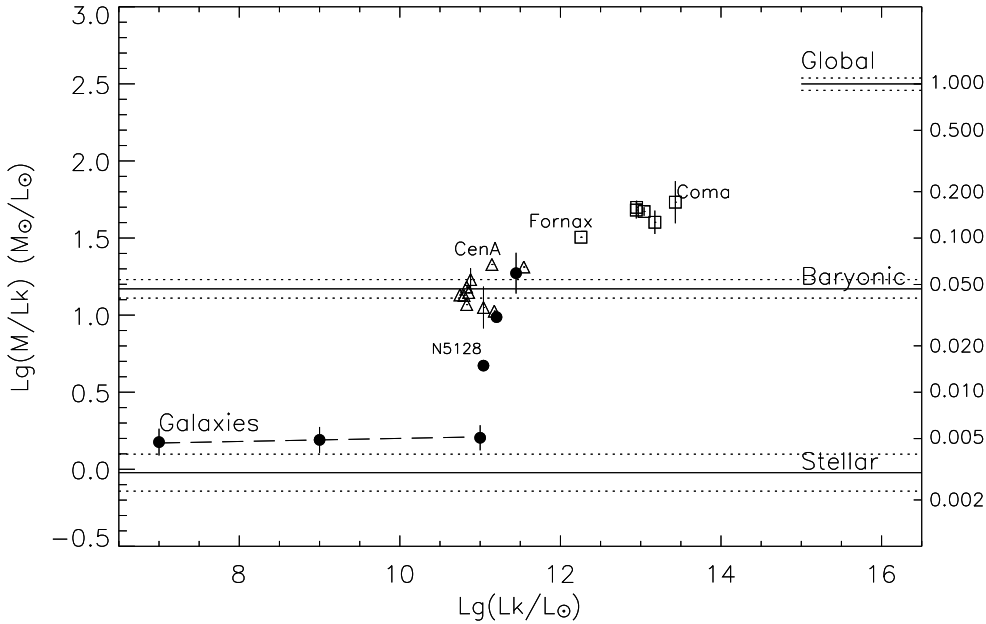}}\par
\vspace{5.0cm}
\caption{ Total mass to K-luminosity ratio for galaxies and systems of
galaxies.
       Three horizontal bands correspond to the stellar component
(bottom),
       the baryonic component (middle), and the global value (top)
defined as
       the critical density per mean luminosity density. The right scale
       shows the mean density of matter in units of the critical
density.
      Infrared luminosities of galaxies in the Local Volume }
\label{pic7.ps}
\end{figure}
\end{document}